\nonstopmode

\newcommand{\XeFn}{XeF$_n$}
\newcommand{\XeFt}{XeF$_2$}
\newcommand{\XeFf}{XeF$_4$}
\newcommand{\XeFs}{XeF$_6$}
\newcommand{\Ft}{F$_2$}
\newcommand{\U}[1]{\,{\rm #1}}
\newcommand{\I}[1]{_{\rm #1}}
\newcommand{\Sum}{\sum\limits}
\newcommand{\ie}{i.e.{}}
\newcommand{\eg}{e.g.{}}
\newcommand{\etal}{{\it et al.{}}}
\newcommand{\mat}[1]{\hbox{\boldmath{$#1$}\unboldmath}}
\newcommand{\unitmatrix}{\mat{\mathbbm{1}}}
\newcommand{\E}[1]{\times 10^{#1}}
\newcommand{\bra}[1]{\left<\right.\!#1\!\left.\right|}
\newcommand{\ket}[1]{\left|\right.\!#1\!\left.\right>}
\newcommand{\brapsi}[1]{\bra{\Psi#1}}
\newcommand{\ketpsi}[1]{\ket{\Psi#1}}
\newcommand{\ketpsiI}{\ketpsi{\I{I}}}
\newcommand{\brapsiI}{\brapsi{\I{I}}}
\newcommand{\psiIHpsiI}{\brapsiI\hat H\ketpsiI}

\documentclass[prb,letterpaper,twocolumn]{revtex4}
\usepackage{graphicx,bbm,amsfonts}
\usepackage[nativepdf,bookmarks,bookmarksopen,bookmarksnumbered,letterpaper,
pdftitle={Impact of Interatomic Electronic Decay Processes on Xe 4d Hole Decay in the Xenon Fluorides},
pdfauthor={Christian Buth, Robin Santra, Lorenz S. Cederbaum},
pdfsubject={Molecular physics, atomic and molecular clusters},
pdfkeywords={interatomic, Auger decay, ICD, ETMD, IAED, ETMD2, ETMD3, interatomic Coulombic decay,
electron trandsfer mediated decay, xenon, fluorides, Xe, F2, XeF2, XeF4, XeF6, ionization,
double ionization, spectrum, electronic resonance, decay width, core hole}]{hyperref}

\begin{document}
\title{Impact of Interatomic Electronic Decay Processes on \\
Xe$\,$4\textit{d}~Hole Decay in the Xenon Fluorides}
\date{19 March 2003}
\author{Christian Buth}
\email[Author to whom all correspondence should be addressed; electronic mail: ]{Christian.Buth@web.de}
\altaffiliation[Present address: ]{Max-Planck-Institut f\"ur
Physik komplexer Systeme, N\"othnitzer Stra\ss{}e~38,
01187~Dresden, Germany}
\author{Robin Santra}
\altaffiliation[Present address: ]{JILA, University of Colorado, Boulder, 
CO 80309-0440}
\author{Lorenz S. Cederbaum}
\affiliation{Theoretische Chemie, Physikalisch-Chemisches
Institut, Ruprecht-Karls-Universit\"at Heidelberg, Im Neuenheimer
Feld~229, 69120~Heidelberg, Germany}

\begin{abstract}
A hole in a $4d$~orbital of atomic xenon relaxes through Auger decay
after a lifetime of~$3 \U{fs}$. Adding electronegative fluorine ligands to 
form xenon fluoride molecules, results in withdrawal of valence-electron
density from Xe. Thus, within the one-center picture of Auger decay,
a lowered Xe$\,4d$~Auger width would be expected, in contradiction,
however, with experiment. Employing extensive \emph{ab initio} calculations 
within the framework of many-body Green's functions, we determine 
all available decay channels in~\XeFn\ and characterize these 
channels by means of a two-hole population analysis. We derive a
relation between two-hole population numbers and partial Auger
widths. On this basis, interatomic electronic decay processes are
demonstrated to be so strong in the xenon fluorides that they
overcompensate the reduction in intra-atomic Auger width and lead
to the experimentally observed trend. The nature of the relevant 
processes is discussed. These processes presumably underlie Auger
decay in a variety of systems.
\end{abstract}
\maketitle

\section{Introduction}
\label{sec:intro}

The Auger effect~\cite{Auger:-23,Burhop:AE-80,Thompson:AE-85,
Bambynek:XR-72,Bambynek:XR-74} provides a magnificent means to
study atoms, molecules, and surfaces. It is caused by a special
type of electronic resonance, i.e., the decay by electron emission
of core-ionized atoms, molecules, or solids. Since its discovery it has 
received much attention, because it is a fundamental process
yielding deep insights into complex many-body effects in matter.
Moreover, the Auger effect has proven to be useful in many experimental 
situations~\cite{Wormeester:IA-91,Burhop:AE-80,Thompson:AE-85,
Bambynek:XR-72,Bambynek:XR-74} and can be used as an analytical tool. 
Therefore, an in-depth understanding of the Auger effect is important both
for fundamental and practical reasons. 

A particularly interesting situation arises when systems consisting 
of more than one atom are considered. Can the Auger decay rate of
a core hole in an atom be influenced---or maybe even adjusted at 
will---by modifying the chemical environment of that atom? If yes,
what are the underlying mechanisms? Early experiments on gas-phase
molecules~\cite{Shaw:CE-72} and ionic solids~\cite{Friedman:CE-72}
hinted at a simple correlation between the binding energy of a core
electron and the Auger width of the associated core hole. For instance,
withdrawal of valence electrons by electronegative ligands induces
a chemical shift of the core level toward higher binding energies; 
at the same time, a smaller number of valence electrons is available 
for Auger decay, if a local, atomic mechanism is assumed. Hence, 
within the one-center picture of Auger decay~\cite{Coville:ME-91}, 
an increase in binding energy is accompanied by a decrease in Auger width. 

Citrin~\cite{Citrin:IAP-73}, however, analyzed x-ray photoemission 
measurements in a number of metal oxides and halides, and found
convincing evidence for interatomic processes, which can lead to a trend 
in contradiction to the one-center model. Citrin's work stimulated theoretical
work on interatomic Auger decay in crystalline NaCl~\cite{Yafet:IAP-77} and 
NaF.~\cite{Green:IAR-87} Matthew and Komninos~\cite{Matthew:TR-75} were the 
first to present a formal examination of interatomic Auger processes. Using 
strong approximations,~\cite{Wormeester:IA-91} they concluded that these 
transitions have a small impact on the Auger rate, except in low-energy Auger 
processes. In fact, recent theoretical and experimental investigations show 
that in the low-energy case the effect of the chemical bond can be 
dramatic.~\cite{Tarantelli:2h-91,Tarantelli:FI-93,Gottfried:FI-96,
Carroll:PE-02,Thomas:ANL-02}

\begin{figure}
  \begin{center}
    \includegraphics[width=\hsize,clip]{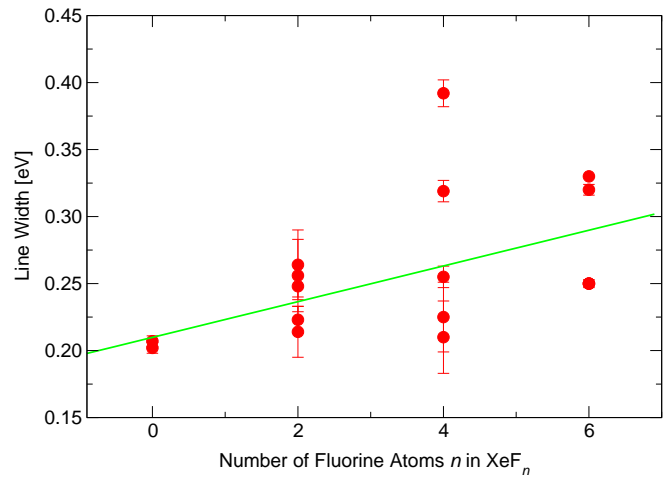}
    \caption{Experimental widths of the Xe$\,4d$~lines in~Xe,
	     \XeFt, \XeFf, and \XeFs~\cite{Cutler:Xe-91}. 
             The figure~\cite{thomas:pri-02} suggests an 
             average increase of the linewidth with an
	     increasing number of fluorine atoms. Note that 
             due to spin-orbit coupling and ligand-field effects, 
             the Xe$\,4d$~orbitals are not all equivalent.}
    \label{fig:Gamma_XeFx}
  \end{center}
\end{figure}

In this work, we argue that interatomic electronic decay processes can
have an even greater impact than previously assumed. We investigate as 
an example Auger decay of a Xe$\,4d$~hole in Xe, \XeFt, \XeFf, and \XeFs.
It is demonstrated that the Auger width in the series \XeFn, $n = 0,2,4,6$,
would decrease drastically if only intra-atomic decay played a role. By
contrast, the experimentally observed linewidths~\cite{Cutler:Xe-91}
display---in clear contradiction to the one-center picture mentioned
previously---a monotonic increase from Xe to \XeFs. This is illustrated
in Fig.~\ref{fig:Gamma_XeFx}.~\cite{thomas:pri-02} We show that interatomic 
Auger processes are responsible for the discrepancy. Vibrational
broadening, which could cause a similar effect, is very likely to be 
insignificant.~\cite{Cutler:Xe-91} 
Single ionization spectra of the xenon fluorides, which we have calculated 
recently,~\cite{Buth:NH-02,Buth:-03} provided a first theoretical indication 
of an entirely electronic mechanism.

The xenon fluorides constitute, from our point of view, an ideal example,
because of the relatively weak bonding between the central noble-gas atom
and the fluorine ligands. This allows us to identify various interatomic
decay processes and to distinguish them from the intra-atomic one. 
In particular, the purely interatomic/intermolecular electronic decay 
mechanisms that govern the inner-valence physics in weakly bound clusters
of atoms or molecules~\cite{Cederbaum:GI-97,Santra:ICD-00,Zobeley:HE-98,
Zobeley:ED-01,Santra:ICD-01,Santra:ED-01,Marburger:EE-03} provide a conceptual
framework for our analysis. Therefore, a short summary of these processes
is given in Sec.~\ref{sec:decayproc}.

The nature of Auger decay in a molecule can be assessed most clearly
by characterizing the available decay channels. To this end, 
Sec.~\ref{sec:compmeth} introduces a many-body Green's function method---the 
algebraic diagrammatic construction scheme~\cite{Schirmer:PP-84,
Gottfried:FI-96}---for calculating double ionization spectra of 
molecules at a highly correlated level. 
A two-hole population analysis~\cite{Tarantelli:2h-91,Tarantelli:RD-92}  
is discussed that serves as an important tool for determining the 
localization properties of the two holes in an Auger decay channel.
Using Wigner-Weisskopf theory,~\cite{Sakurai:MQM-94,Weisskopf:-30} we 
derive a relationship between final-state population numbers and 
partial Auger widths. The calculated double ionization spectra are analyzed 
in Sec.~\ref{sec:results}, and the relevance of interatomic decay
processes for the decay of Xe$\,4d$~vacancies is elucidated.
Concluding remarks are the subject of Sec.~\ref{sec:conclusion}.

\section{Electronic Decay Processes in Weakly Bound Systems}
\label{sec:decayproc}

In this section, we review the knowledge obtained in the study of
the electronic decay of inner-valence ionized clusters. The decay 
processes discovered in clusters, and the associated terminology we 
use in order to emphasize the nature of the underlying mechanisms, 
allow a succinct classification of the Auger processes in \XeFn.
We mention that the decay processes discussed in the following for singly 
ionized systems can be generalized to describe the decay of inner-valence 
vacancies of multiply ionized clusters.~\cite{Zobeley:HE-98,Santra:CE-03}

\begin{figure*}
  \begin{center}
   \includegraphics[width=\hsize,clip]{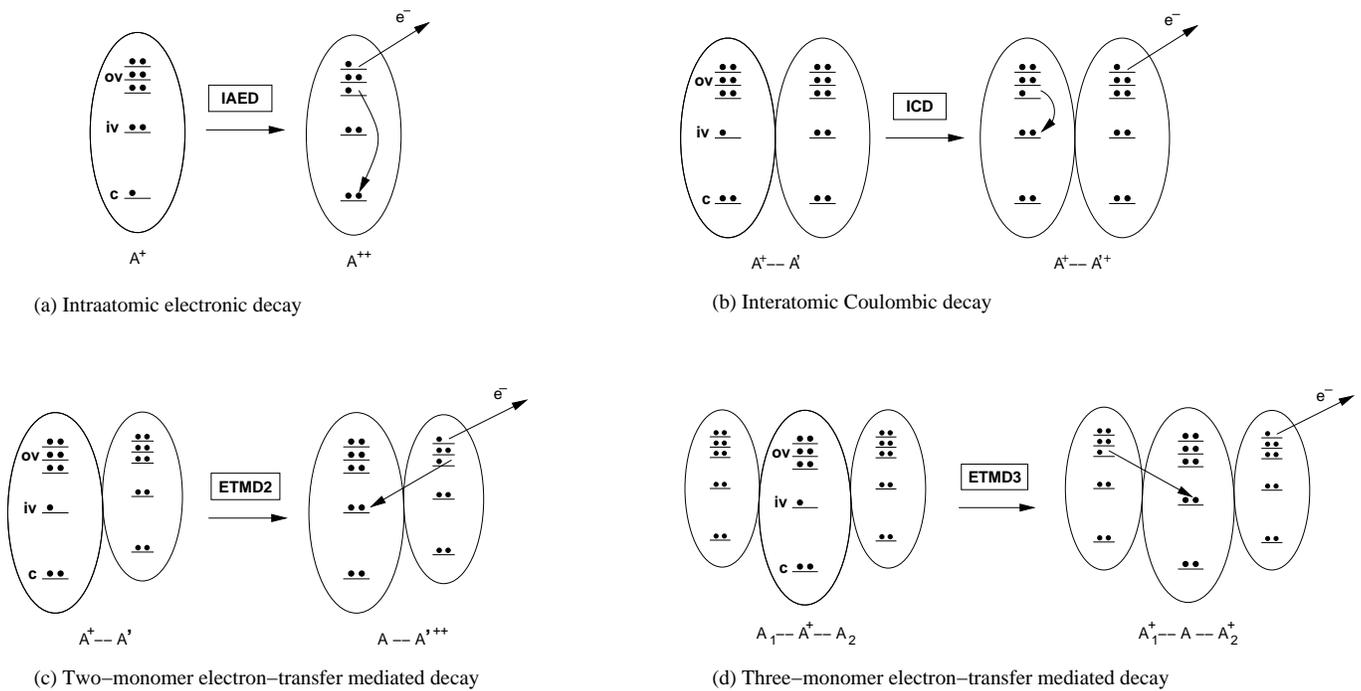}
  \end{center}
  \caption{Principles of electronic decay processes in weakly bound 
           systems: (a)~intra-atomic electronic decay~(IAED), 
           (b)~interatomic Coulombic decay~(ICD), (c)~two-monomer 
           electron-transfer mediated decay~(ETMD2), (d)~three-monomer 
           electron-transfer mediated decay~(ETMD3). Common to all 
           electronic decay processes is that the initial vacancy is 
           filled by a valence electron from the same or a neighboring 
           atom. The excess energy is transferred to a second valence 
           electron of the same or a neighboring atom. Emission of this 
           electron results in distinct dicationic charge distributions 
           that allow one to characterize the underlying decay process.}	 
  \label{fig:decays}
\end{figure*}

\emph{Intra-atomic electronic decay}~(IAED) is usually the most prominent
electronic decay process of core-ionized atoms embedded in molecules or 
clusters. IAED is, of course, the only Auger process that can take place
in an isolated atom. In atoms, this mechanism is well 
understood:~\cite{Auger:-23,Burhop:AE-80,Thompson:AE-85,Bambynek:XR-72,
Bambynek:XR-74} The initial core hole is filled by a (valence) 
electron and the excess energy is
transferred via Coulomb interaction to a second (valence) electron, 
which is emitted subsequently~[Fig.~\ref{fig:decays}(a)]. In molecules or 
clusters, intra-atomic decay is not the only decay process, because 
other processes can occur that involve neighboring atoms or molecules.

If one removes an inner-valence electron from an isolated atom or
molecule, the ionized system cannot, in general, decay by electron 
emission, for it lies energetically below the double ionization threshold. 
The situation changes if the monomer is part of a cluster. 

The initial inner-valence vacancy is filled by a valence electron 
of the same monomer and the excess energy is transferred to a second 
valence electron of a neighboring monomer. This electron is emitted 
subsequently. See Fig.~\ref{fig:decays}(b) for a schematic representation. 
The decay process is termed \emph{interatomic} or \emph{intermolecular
Coulombic decay}~(ICD) for clusters of atoms or clusters of
molecules, respectively. It was identified theoretically in several 
weakly bound clusters: (HF)$_n$ clusters,~\cite{Santra:ICD-01,
Cederbaum:GI-97,Zobeley:HE-98,Santra:ED-99} the HF(H$_2$O)$_2$
cluster,~\cite{Zobeley:ICD-99} (H$_2$O)$_n$
clusters,~\cite{Cederbaum:GI-97} Ne$_n$
clusters,~\cite{Santra:ICD-00,Santra:ED-01,Scheit:TD-03} and
the NeAr dimer.~\cite{Zobeley:ED-01} Further we would like to point out
studies of electronic decay after inner-valence ionization of 
CN$^-$~\cite{Santra:IVI-00} and of fluorinated carbanions and their 
acids.~\cite{Muller:CED-02} 

ICD of inner-valence holes in clusters is ultrafast, typically taking place 
on a time scale of the order of $10$~fs (see Ref.~\onlinecite{Santra:NH-02} 
for a review). Experimental evidence for ICD has been found recently
in neon clusters.~\cite{Marburger:EE-03} In the context of this paper, it 
is especially important to note that ICD is highly sensitive to cluster
size: The ICD~width in Ne$_n$~clusters, for instance, increases strongly 
with~$n$, when a central neon atom carries the initial inner-valence vacancy 
and the surrounding atoms converge with~$n$ to the shape of the first 
coordination sphere in solid neon.~\cite{Santra:ED-01} Clearly, a larger 
number of nearest neighbors translates into a larger number of dicationic 
decay channels available for~ICD.

The ionization spectra of neon and argon are quantitatively very different
from one another (Ref.~\onlinecite{Zobeley:ED-01}, and references therein). 
Neon has significantly higher lying single~(IP) and double ionization 
potentials~(DIP) compared to argon. The inner-valence Ne$\,2s$~IP is larger 
than the Ne$^{-1}$Ar$^{-1}$~DIP in the NeAr dimer.~\cite{Zobeley:ED-01} 
Therefore, an initial Ne$\,2s$~vacancy can lead to~ICD. However, the 
Ne$\,2s$~vacancy in NeAr can decay in another way, too, because the 
Ne$\,2s$~IP is also larger than the NeAr$^{-2}$~DIP. In this case, a valence 
electron from argon drops into the Ne$\,2s$ hole and the excess energy is 
transferred to another valence electron of argon. Upon electron emission, 
the dimer is left in a NeAr$^{-2}$ state. This process is called 
\emph{electron-transfer mediated decay}~(ETMD).~\cite{Zobeley:ED-01}
A schematic representation of the ETMD process is shown in 
Fig.~\ref{fig:decays}(c). In Ref.~\onlinecite{Zobeley:ED-01} it is shown that 
the contribution of ETMD to the total electronic decay width of an initial
Ne$\,2s$~vacancy in NeAr is appreciably smaller than the contribution of~ICD. 
In light of the vastly increasing ICD width in neon clusters,~\cite{Santra:ED-01} 
we expect a similar size effect for~ETMD, too. In the ETMD described, only 
two atoms are involved. It is called \emph{two-monomer ETMD}~(ETMD2).

A \emph{three-monomer ETMD}~(ETMD3) has been suggested to occur for an initial 
Ne$\,2s$~vacancy in NeAr$_2$.~\cite{Zobeley:ED-01} A schematic representation 
of this (proposed) process is given in Fig.~\ref{fig:decays}(d). A valence 
electron of one of the two argon atoms is transferred into the initial 
Ne$\,2s$~hole. Energy is released, and a valence electron is ejected from the 
second argon atom, leaving the trimer in a NeAr$^{-1}$Ar$^{-1}$~state.

We would like to study a new example for the surprising impact of
the ideas presented so far. To this end, we examine the decay of
Xe$\,4d$ vacancies in~XeF$_n$ and demonstrate how interatomic processes 
lead to an increase in Auger width. In order to pursue this issue, we need 
to introduce a few computational techniques and useful relations in the ensuing 
section.

\section{Computational Methods and Basic Relations}
\label{sec:compmeth}
\subsection{Algebraic Diagrammatic Construction and Two-Hole
Population Analysis}
\label{sec:doublepop}

Many-body Green's functions are ideal tools for investigating excitation
spectra of molecular matter at various stages of 
ionization.~\cite{Cederbaum:TA-77,Cederbaum:GF-98} The two-particle 
Green's function, specifically, provides direct access to double ionization
spectra and associated eigenstates. Algebraic diagrammatic 
construction~(ADC) is an advanced perturbation-theoretical approximation 
scheme for evaluating many-body propagators.~\cite{Schirmer:PP-82,
Schirmer:GF-83,Schirmer:PP-84,Schirmer:SE-89} 
ADC($n$) sums all Feynman diagrams up to $n$th order as well as 
certain classes of diagrams up to infinite order. There is no Dyson 
equation for the two-particle Green's function.~\cite{Brand:ET-96,Brand:FO-98,
Brand:TE-01} The two-particle ADC evaluates the two-particle propagator 
directly in terms of a Hermitian eigenvalue problem.~\cite{Schirmer:PP-84} 
The program~\cite{Tarantelli:BP-03} we use to calculate double ionization 
potentials implements the two-particle ADC(2) scheme.~\cite{Schirmer:PP-84}

The two-hole population analysis, utilized in combination with the 
ADC(2)~program,~\cite{Tarantelli:BP-03} is a means to reveal the spatial
localization of the two holes resulting from double 
ionization.~\cite{Tarantelli:2h-91,Tarantelli:RD-92} The analysis is
carried out in a Mulliken-type~\cite{Szabo:MQC-82} fashion by determining the 
atomic two-hole ($2h$) contributions to the dicationic eigenstates. 
To this end, a dicationic eigenstate in the ADC(2) scheme is first expanded 
in terms of $2h$~configurations in molecular orbital basis,
\begin{equation}
  \label{eq:dipExp}
  \ket{\Psi_n^{N-2}} = \sum_{ij} (\vec X_n)_{ij} \ket{ij} + \cdots ,
\end{equation}
where $\vec X_n$~is the 2$h$ part of the ADC~eigenvector of the
$n$th~dicationic state $\ket{\Psi_n^{N-2}}$. The symbol $\ket{ij}$ denotes 
a spin-adapted $2h$ configuration with holes in molecular (spin) orbitals 
$i$ and $j$, both of which are occupied in the Hartree-Fock ground state of 
the $N$-electron system. Note that $ij$ runs only over distinguishable 
$2h$ configurations. 
Each $2h$ configuration in molecular orbital basis, $\ket{ij}$, 
can now be expanded~\cite{Tarantelli:2h-91,Tarantelli:RD-92} in terms of
2$h$ configurations~$\ket{\mu\nu}$ deriving from the atomic orbital
basis
\begin{equation}
  \label{eq:dipAO}
  \ket{ij} = \sum_{\mu\nu} U_{\mu\nu,ij} \, \ket{\mu\nu} \; .
\end{equation}
Using this transformation, the total 2$h$ weight~$\vec X_n^{\dagger} \vec X_n$
(2$h$ pole strength) of the $n$th dicationic eigenstate can be written 
as~\cite{Tarantelli:2h-91,Tarantelli:RD-92}
\begin{equation}
  \label{eq:2hpoppole}
  \vec X_n^{\dagger} \vec X_n = \vec Y_n^{\dagger} \mat O
  \vec Y_n = \sum_{\mu\nu} \Bigl[ \underbrace{Y_{\mu\nu,n} \sum_{\rho\sigma}
  \mat O_{\mu\nu,\rho\sigma} Y_{\rho\sigma,n}}_{Q_{\mu\nu,n}} \Bigr] \; ,
\end{equation}
with eigenvector expansion coefficients~$\vec Y_n := \mat U \vec X_n$ 
in atomic orbital basis. $\mat O$ represents the overlap matrix between atomic
$2h$ configurations, which satisfies
$\mat U^{\dagger} \mat O \mat U = \unitmatrix$.

On choosing suitable sets of atomic basis functions~$A$ and $B$, one
obtains population numbers~\cite{Tarantelli:2h-91,
Tarantelli:RD-92} $Q_{AB,n} = \Sum_{\mu \in A \atop \nu \in B}
Q_{\mu\nu,n}$. Frequently, as is done in this work, $A$~and~$B$
represent the atomic basis functions associated with two not necessarily 
distinct atoms within a molecule. In this sense, $A$~and~$B$ can be identified
with individual atoms. This provides a clear picture of the localization
properties of the two holes.

\subsection{Relation between Two-Hole Population Numbers
and Partial Decay Widths}
\label{sec:widths}

A useful approximation to the decay width of a resonance is
given within the framework of Wigner-Weisskopf 
theory,~\cite{Sakurai:MQM-94,Weisskopf:-30} which is founded on 
time-dependent perturbation theory. Applications of Wigner-Weisskopf
theory to the electronic decay of inner-valence-ionized clusters can 
be found in Refs.~\onlinecite{Santra:ED-01} and \onlinecite{Zobeley:ED-01}. 
There, single Hartree-Fock determinants were employed to describe both 
initial and final states. For more complicated systems like \XeFn\ and on 
the energy scale of core levels, however, the final states in particular 
require a more sophisticated treatment.

Let the initial state~$\ket{\Psi\I{I}^{N-1}}$ be a discrete, singly ionized
state approximating the core-hole resonance and a final state be a doubly 
ionized state~$\ket{\Psi_n^{N-2}}$ together with a decay electron of 
momentum~$\vec k$. For simplicity we employ the sudden 
approximation (Ref.~\onlinecite{Hedin:SA-02}, and references
therein) and express such a final state by the antisymmetrized 
product~${\hat c}_{\vec k}^{\dagger} \ket{\Psi_n^{N-2}}$, where
${\hat c}_{\vec k}^{\dagger}$ is a fermionic creation 
operator.~\cite{Fetter:MP-71} The decay width is then given by
\begin{equation}
  \label{eq:WWdecaywidth}
  \Gamma = 2\pi \sum_n \sum_{{\vec k}} 
  |\bra{\Psi_n^{N-2}}{\hat c}_{\vec k} {\hat H} \ket{\Psi\I{I}^{N-1}}|^2 
  \, \delta(E_{n {\vec k}} - E\I{I}) \; .
\end{equation}
This equation contains three types of matrix elements: transition matrix 
element $\bra{\Psi_n^{N-2}}{\hat c}_{\vec k} {\hat H} \ket{\Psi\I{I}^{N-1}}$;
initial-state energy $E\I{I} := \psiIHpsiI$; and final-state energy
$E_{n {\vec k}} := \bra{\Psi_n^{N-2}}{\hat c}_{\vec k} {\hat H}
{\hat c}_{\vec k}^{\dagger} \ket{\Psi_n^{N-2}}$. The operator ${\hat H}$
is the Hamiltonian in fixed-nuclei approximation. The $\delta$ function in
Eq.~(\ref{eq:WWdecaywidth}) ensures that the contribution to
the total decay width of only those accessible final states is
summed that conserve energy in the decay.

Using Eq.~(\ref{eq:dipExp}), we obtain
\begin{equation}
  \label{eq:transexp}
  \bra{\Psi_n^{N-2}}{\hat c}_{\vec k} {\hat H} \ket{\Psi\I{I}^{N-1}} 
  = \sum_{ij} (\vec X_n^{\dagger})_{ij} \bra{ij} {\hat c}_{\vec k} 
  \hat H \ket{\Psi\I{I}^{N-1}} + \cdots
\end{equation}
for the transition matrix elements in Eq.~(\ref{eq:WWdecaywidth}).
Let us assume that the initial state can be approximated by a single 
determinant~$\ket{\Psi\I{I}^{N-1}} = \hat c_l \ket{\Phi_0^{N}}$, 
where~$l$ indicates a core orbital and $\ket{\Phi_0^{N}}$ the $N$-electron
Hartree-Fock ground state. Let us further exploit the fact that excitations of
core orbitals are, to a good approximation, negligible in all expansion 
terms of the dicationic final state $\ket{\Psi_n^{N-2}}$. The terms 
indicated in Eq.~(\ref{eq:transexp}) with ``$\cdots$'' then vanish exactly 
due to the Slater-Condon rules.~\cite{Szabo:MQC-82}

To proceed, we express the dicationic electronic configurations in terms of
atomic $2h$ vectors $\ket{\mu\nu}$. The configurations~$\ket{ij}$ appearing 
in Eq.~(\ref{eq:transexp}) form an orthonormalized set of functions, and by using 
the relation
\begin{equation}
  \label{eq:orthexp}
  \ket{ij} = \sum_{\mu\nu,\rho\sigma} 
  (\mat O^{\frac{1}{2}})_{\mu\nu,\rho\sigma}
  \, U_{\rho\sigma,ij} \, \ket{\mu\nu}  \; ,
\end{equation}
we introduce an \emph{orthonormalized} set of configurations~$\ket{\mu\nu}$ 
referring to atomic orbitals. Notice that the atomic configurations 
in Eq.~(\ref{eq:dipAO}) are not orthonormalized. Inserting 
Eq.~(\ref{eq:orthexp}) into Eq.~(\ref{eq:transexp}) leads to
\begin{equation}
  \label{eq:transme}
  \bra{\Psi_n^{N-2}}{\hat c}_{\vec k} {\hat H} \ket{\Psi\I{I}^{N-1}} =
  \sum_{\mu\nu} (\vec Y_{n}^{\prime \, \dagger})_{\mu\nu}
  \bra{\mu\nu} {\hat c}_{\vec k} \hat H \ket{\Psi\I{I}^{N-1}} \; ,
\end{equation}
where~$\vec Y_{n}^{\prime} := \mat O^{\frac{1}{2}} \mat U \vec
X_n$. This equation relates the transition matrix element to a
superposition of atomic-like quantities.

We now insert the transition matrix element~(\ref{eq:transme})
into the basic Eq.~(\ref{eq:WWdecaywidth}) for the decay width, 
integrate over the momentum~$\vec k$ of the emitted electron, and
neglect interference terms (cross terms). The result reads
\renewcommand{\arraystretch}{1.5}%
\begin{equation}
  \begin{array}{rcl}
    \Gamma &=& \Sum_n \Gamma_n \; , \\
    \Gamma_n &=& \Sum_{\mu\nu} |T_{\mu\nu}|^2 \, |Y_{\mu\nu,n}^{\prime}|^2 \; .
  \end{array}
\end{equation}
\renewcommand{\arraystretch}{1}%
The partial width $\Gamma_n$ associated with the final dicationic
state~$\ket{\Psi_n^{N-2}}$ is expressed as a sum over atomic $2h$
configurations. Each contributing term consists of a state-specific 
factor~$|Y_{\mu\nu,n}^{\prime}|^2$, which we have calculated
earlier. The other factor,
\begin{equation}
  |T_{\mu\nu}|^2 = 2\pi \sum_{\vec k} |\bra{\mu\nu} {\hat c}_{\vec k} 
  \hat H \ket{\Psi\I{I}^{N-1}}|^2 \, \delta(E_{n {\vec k}} - E\I{I}) \; ,
\end{equation}
depends on the dicationic state~$\ket{\Psi_n^{N-2}}$ only via the
$\delta$ function and is termed transition strength. Since we consider the 
decay of a core level, implying that the energy of the emitted electron is 
relatively high, this dependence on $n$ is weak and can be neglected to a 
good approximation.

The state-specific quantities~$|Y_{\mu\nu,n}^{\prime}|^2$ are closely
related to the occupation numbers discussed in Sec.~\ref{sec:doublepop}. 
In principle, they could be used as an alternative definition of occupation 
numbers in the spirit of L\"owdin's population analysis.~\cite{Szabo:MQC-82}
In particular, if the atomic basis functions are well localized, each on its
atom, we may put~$|Y_{\mu\nu,n}^{\prime}|^2 = Q_{\mu\nu,n}$ and obtain
\begin{equation}
  \label{eq:ICDpop}
  \Gamma_n = \Sum_{\mu\nu} |T_{\mu\nu}|^2 \, Q_{\mu\nu,n} \; .
\end{equation}
For a given molecule the quantities~$|T_{\mu\nu}|^2$ are universal, and
the only information that is state-specific is contained in the occupation
numbers~$Q_{\mu\nu,n}$. We are able to compute the occupation numbers
of the molecules studied in this work, and hence Eq.~(\ref{eq:ICDpop})
is already a useful tool of analysis.

We note, however, that many elements~$|T_{\mu\nu}|^2$ can be of
similar magnitude, \eg, if~$\mu$ denotes a $p$-type atomic orbital
on atom~$A$ and~$\nu$ a $p$-type atomic orbital on atom~$B$. We therefore 
group together in Eq.~(\ref{eq:ICDpop}) the corresponding contributions: 
$\Gamma_n = |T_{AB}|^2 \Sum_{\mu \in A \atop \nu \in B}
Q_{\mu\nu,n} + \cdots$. Since the~$|T_{\mu\nu}|^2$ grouped together
are not identical, the average quantity~$|T_{AB}|^2$ will depend somewhat 
on the dicationic state index~$n$. Computing~$\Gamma = \Sum_n \Gamma_n$ 
amounts to averaging over the slightly varying~$|T_{AB}|^2$ and thus leads 
to a more stable result.

Highly localized dicationic final-state holes, for example those that
result from ICD~in weakly bound clusters, lead to similar contributions 
to the overall decay width of the initially ionized state. By classifying 
the atomic pairs~$(\mu\nu)$ by their localization characteristics, we can 
form sets of atomic orbitals where~$|T_{AB}|^2$ is only a slightly varying 
quantity. Let us denote with~$A$ the atom on which the initial core hole is
localized. Then, if $\mu$~and~$\nu$ denote atomic orbitals on~$A$,
$|T_{\mu\nu}|^2$ obviously corresponds to IAED (Sec.~\ref{sec:decayproc}).
Analogously, if $\mu$~belongs to atom~$A$ and $\nu$~to
atom~$B$, we are dealing with~ICD, and if neither~$\mu$ nor~$\nu$
belong to~$A$, $|T_{\mu\nu}|^2$~describes either ETMD2 or ETMD3,
depending on whether $\mu$~and~$\nu$ belong to the same atom~$B$ or
to two atoms~$B$ and~$C$. Thus, the \emph{minimal} dissection of~$\Gamma$ 
into its various contributions, in the spirit of the above discussion, reads
\begin{equation}
  \label{eq:partial}
   \Gamma = \Gamma\I{IAED} + \Gamma\I{ICD}
            + \Gamma\I{ETMD2} + \Gamma\I{ETMD3} \; ,
\end{equation}
where
\renewcommand{\arraystretch}{1.5}%
\begin{equation}
  \label{eq:defs1}
  \begin{array}{rcl}
    \Gamma\I{IAED}  &=& |T\I{IAED}|^2  \; Q\I{IAED}  \; , \\
    \Gamma\I{ICD}   &=& |T\I{ICD}|^2   \; Q\I{ICD}   \; , \\
    \Gamma\I{ETMD2} &=& |T\I{ETMD2}|^2 \; Q\I{ETMD2} \; , \\
    \Gamma\I{ETMD3} &=& |T\I{ETMD3}|^2 \; Q\I{ETMD3} \; , 
  \end{array} 
\end{equation}
\renewcommand{\arraystretch}{1}%
and
\renewcommand{\arraystretch}{1.5}%
\begin{equation}
  \label{eq:defs2}
  \begin{array}{rcl}
    Q\I{IAED}  &=& \Sum_n \Sum_{\mu,            \nu \in A} Q_{\mu\nu,n} \; , \\
    Q\I{ICD}   &=& \Sum_n \Sum_{\mu \in A \atop \nu \in B} Q_{\mu\nu,n} \; , \\
    Q\I{ETMD2} &=& \Sum_n \Sum_{\mu,            \nu \in B} Q_{\mu\nu,n} \; , \\
    Q\I{ETMD3} &=& \Sum_n \Sum_{\mu \in B \atop \nu \in C} Q_{\mu\nu,n} \; .
  \end{array}
\end{equation}
\renewcommand{\arraystretch}{1}%
If necessary, further dissections of~$\Gamma$ into more detailed
contributions are readily done. For instance, grouping together
separately the contributions of~$\mu$ and~$\nu$, both being atomic
orbitals of $s$ type, both of $p$ type, or one of $s$ and one of
$p$ type, etc. A distinction between the decay into singlet and
triplet final dicationic states is also possible. Finally, we
mention that the individual transition strengths, $|T\I{ICD}|^2$,
etc., may be similar in a series of related molecules.

\section{Results and Discussion}
\label{sec:results}
\subsection{Double Ionization Spectra}

\begin{figure*}
  \begin{center}
    \includegraphics[width=\hsize,clip]{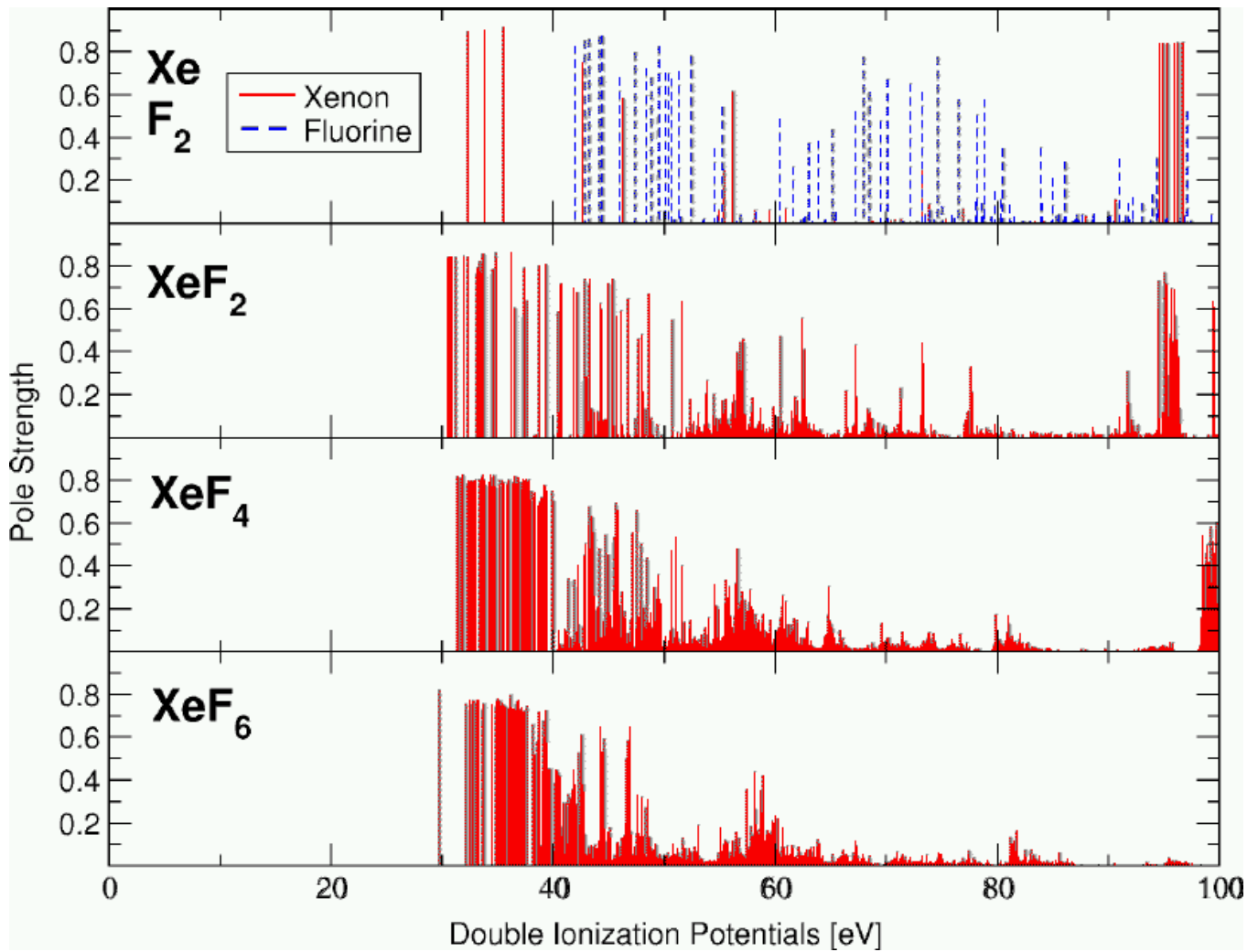}
    \caption{(Color) Double ionization spectra of~Xe, \Ft, \XeFt,
             \XeFf, and \XeFs.  The two-particle ADC(2) $2h$ pole
             strength is plotted on the ordinate to characterize
             how well the dicationic final states are described by
             $2h$ configurations.}
    \label{fig:XeFx_dip}
  \end{center}
\end{figure*}

The computed double ionization spectra for the xenon atom, the fluorine 
molecule, and the series~XeF$_n$, $n=2,4,6$, are shown in 
Fig.~\ref{fig:XeFx_dip}. The molecular ground-state geometries used in the 
calculations are described in Refs.~\onlinecite{Buth:NH-02} and \onlinecite{Buth:-03}.
It has to be stressed, however, that the double ionization spectrum of~\XeFs, 
unlike the single ionization spectrum in Refs.~\onlinecite{Buth:NH-02} and
\onlinecite{Buth:-03}, has been calculated assuming a geometry of $O_h$~symmetry, 
instead of $C_{3v}$~symmetry, in order to make the calculation feasible with 
the available computer resources. We also refer the reader to 
Refs.~\onlinecite{Buth:NH-02} and \onlinecite{Buth:-03} for details regarding 
the Gaussian basis sets employed. 

The ADC $2h$~pole strength plotted in Fig.~\ref{fig:XeFx_dip} characterizes 
how well the dicationic states are described by $2h$~configurations. The 
deviation of the $2h$~pole strength from unity yields the total contribution 
of three-hole--one-particle configurations to the state in question. Our focus
in this work is on the energy regime below the Xe$\,4d$~single ionization
threshold, which ranges from almost $68 \U{eV}$ in atomic Xe to a little less 
than $76 \U{eV}$ in XeF$_6$.~\cite{Cutler:Xe-91,Buth:NH-02,Buth:-03}
The double ionization spectra below~$80 \U{eV}$ in Fig.~\ref{fig:XeFx_dip}
comprise ionization out of the outer and inner valence of Xe, \Ft, and 
XeF$_n$. Since mean-field and correlation effects, as revealed by the ADC 
scheme, dominate in the valence region, the spectra have not been corrected 
for relativistic effects, which are expected to be negligible.

Double ionization from the outer valence of the xenon atom takes
place between~$30$ and $40 \U{eV}$, and three distinct DIPs
are found in Fig.~\ref{fig:XeFx_dip}. The spectra of the series~XeF$_n$,
$n=2,4,6$, display in this region increasingly dense lying DIPs, which is
attributed to a growing number of outer-valence states due to a higher 
delocalization of the two final-state holes. The double ionization spectrum 
of~\Ft~in Fig.~\ref{fig:XeFx_dip} shows outer-valence ionization of~\Ft~in
the range between~$40$ and $55 \U{eV}$. With an increasing number of
fluorine ligands, the general trend in the spectra, especially at DIPs 
above~$40 \U{eV}$, is a decrease of the validity of the two-hole 
description of dicationic states of XeF$_n$. As all fluorine atoms 
in~XeF$_n$ are equivalent (the double ionization spectrum of~\XeFs~is 
obtained in~$O_h$ symmetry), a dicationic state with a hole on xenon 
and a hole on a fluorine atom is in fact a linear combination of 
configurations involving a hole on xenon and a hole on one of the $n$
fluorine atoms in~XeF$_n$. The coefficients of the linear combinations are
determined by the strength of the interaction among the fluorine
ligands, being the weakest in~\XeFt.

The triple ionization threshold of the xenon atom~\cite{Mathur:IX-87} 
is~$66.2 \U{eV}$; for the xenon fluorides, the triple ionization  
threshold is at least that low. Therefore, many of the dicationic states 
we have calculated can decay to tricationic ones. Owing to the discrete 
basis set used, the appearance of a decaying state mimics a discretized 
Lorentzian curve.~\cite{Zobeley:HE-98} As can be seen in 
Fig.~\ref{fig:XeFx_dip}, such curves may indeed be identified for~XeF$_n$
above~$50 \U{eV}$. Interestingly, the Xe$\,4d$~lines in the single ionization 
spectra of xenon and its fluorides~\cite{Cutler:Xe-91,Buth:NH-02,Buth:-03}
are thus above the triple ionization threshold. This indicates that there 
is some probability for the Xe$\,4d$~holes to decay by emitting \emph{two}
electrons. According to a measurement by 
K\"ammerling~\etal,~\cite{Kaemmerling:DA-92} this probability amounts to 
about~$20 \, \%$ in atomic xenon.

When analyzing the~DIPs, one has to keep in mind that essentially
there are four contributions to a dicationic state that determine its
character. The two holes in the dicationic state can be localized
on \emph{one} atom, either Xe or F, or on \emph{two} atoms,
Xe~and F or two different F~atoms. The one-site states have a
large dicationic population number on either xenon (Xe$^{-2}$) or 
a single fluorine (F$^{-2}$), and the two-site states are of
either~Xe$^{-1}$F$^{-1}$ or~F1$^{-1}$F2$^{-1}$ character. For
each dicationic state appearing in the spectra of the xenon
fluorides, the data of the population analysis for equivalent
fluorine atoms are summed up to yield the population numbers. In~\XeFt,
for example, the Xe$^{-1}$F1$^{-1}$ and Xe$^{-1}$F2$^{-1}$ population
numbers are added to give a single Xe$^{-1}$F$^{-1}$~contribution for each 
dicationic state in the spectrum of~\XeFt.

\subsection{One-site Population Numbers}
\label{sec:one-site}

\begin{figure*}
  \begin{center}
    \includegraphics[width=\hsize,clip]{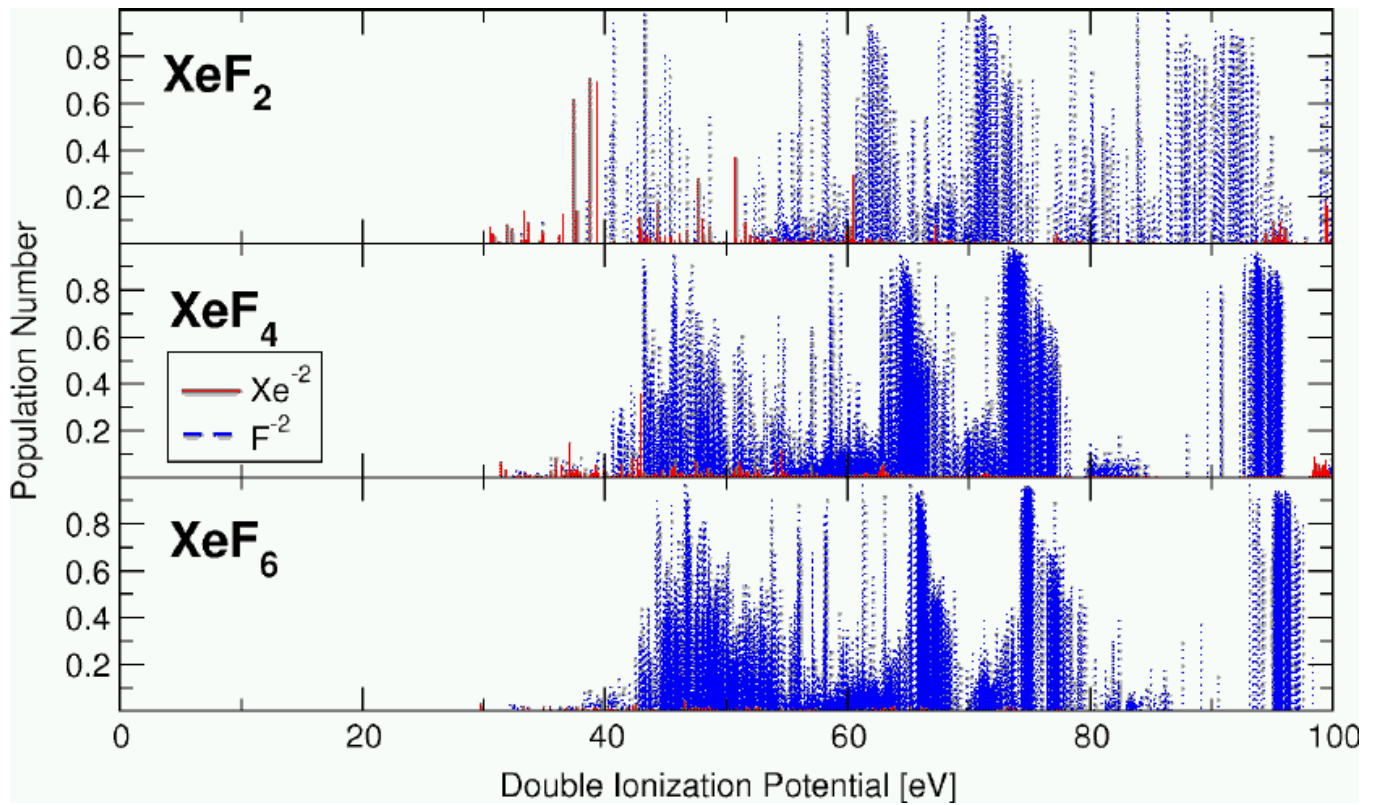}
    \caption{(Color) One-site population numbers of the double
	     ionization spectra of~\XeFt, \XeFf, and \XeFs. Each
	     line shown is related to a dicationic state in the
	     double ionization spectra in Fig.~\ref{fig:XeFx_dip}.
	     The height of the line gives the one-site population
	     number. Note that the contributions on xenon decrease 
             drastically as the number of fluorine atoms increases.}
    \label{fig:dip1}
  \end{center}
\end{figure*}

The one-site population numbers for the xenon fluorides are plotted in
Fig.~\ref{fig:dip1}. For the graphical representation in Fig.~\ref{fig:dip1} and 
in Fig.~\ref{fig:dip2}, the population numbers have been normalized, \ie, 
the sum of the contributions of~Xe$^{-2}$, F$^{-2}$, Xe$^{-1}$F$^{-1}$, and 
F1$^{-1}$F2$^{-1}$~character equals unity for each dicationic eigenstate.
The spectra are compared to study the effect of the 
increasing number of fluorine atoms. The first states in the double 
ionization spectra possess~DIPs of~$\approx 30 \U{eV}$. The major 
contribution of the one-site population~Xe$^{-2}$ is at the lower energy 
part of the spectrum. Obviously, the density of states with a considerable
F$^{-2}$~population increases due to the increasing number of
fluorine atoms. The overall distribution of these states does not
change much in the different compounds. Certain regions are
visible where the states have a high F$^{-2}$~population. These
regions shift slightly to higher~DIPs due to the reduced excess
charge that the individual fluorine atoms acquire from the
central xenon atom in the molecular ground state.~\cite{Buth:NH-02,Buth:-03}

Conversely, the importance of Xe$^{-2}$~contributions to the
dicationic states is extremely reduced due to the withdrawal of 
valence electron density on the xenon atom. In~\XeFs, the
contributions of Xe$^{-2}$~character have nearly vanished.

\subsection{Two-site Population Numbers}

\begin{figure*}
  \begin{center}
    \includegraphics[width=\hsize,clip]{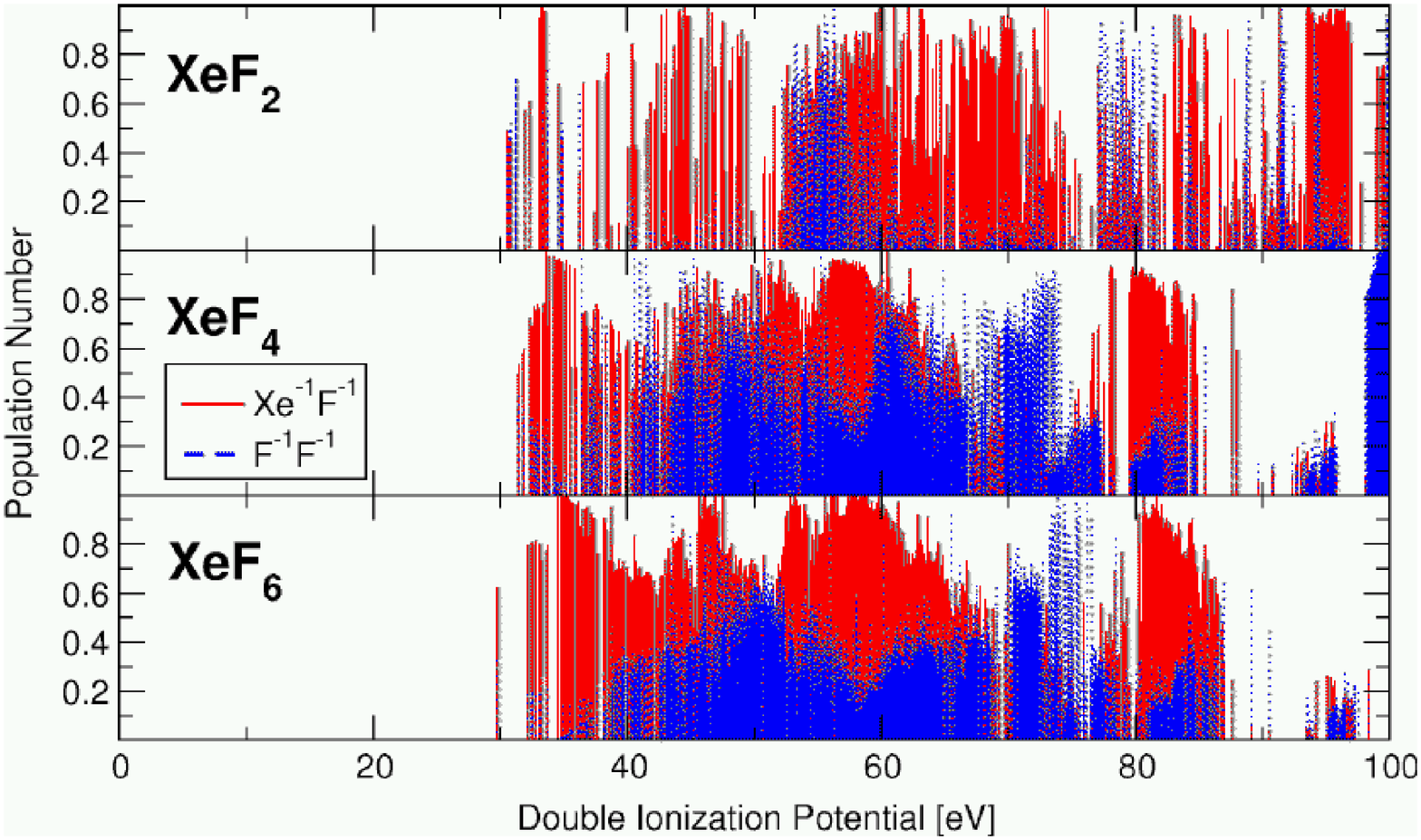}
    \caption{(Color) Two-site population of the double
	     ionization spectra of~\XeFt, \XeFf, and \XeFs.}
    \label{fig:dip2}
  \end{center}
\end{figure*}

The two-site population numbers for the xenon fluorides are
plotted in Fig.~\ref{fig:dip2}. The impact of the increasing
number of fluorine atoms is seen here as well. The states with a
large F1$^{-1}$F2$^{-1}$~population in the spectrum of~\XeFt\ are
clearly separated into distinct groups of lines originating from
F1$\,2p^{-1}\,\,$F2$\,2p^{-1}$, F1$\,2p^{-1}\,\,$F2$\,2s^{-1}$, and
F1$\,2s^{-1}\,\,$F2$\,2s^{-1}$ populations.  This can be concluded
from a simple energy consideration. The IPs for ionization from 
molecular orbitals with F$\,2p$~character are located 
at~$\approx 20 \U{eV}$, those for ionization from molecular orbitals 
with~F$\,2s$ character are located at~$\approx 40 
\U{eV}$.~\cite{Buth:NH-02,Buth:-03} In the spectrum of~\XeFt\ in 
Fig.~\ref{fig:dip2}, the groups are approximately at~40, 60, and 
$80 \U{eV}$. \XeFt\ is a linear molecule, the two fluorine atoms 
being separated by the central xenon atom. Such states with a large 
F1$^{-1}$F2$^{-1}$~population are termed \emph{opposite} 
F1$^{-1}$F2$^{-1}$~states.

The situation changes in~\XeFf\ and \XeFs. There are 
\emph{adjacent} and opposite F1$^{-1}$F2$^{-1}$ states, and the clear
separation between the groups seen in the spectrum of~\XeFt\ is
lost. One reason for this effect is the interaction between
adjacent fluorine atoms, which is stronger than the interaction
between opposite ones. This leads to a splitting of the fluorine
lines, which is also observed in the single ionization
spectra.~\cite{Buth:NH-02,Buth:-03}

The F1--F2~distance in adjacent F1$^{-1}$F2$^{-1}$~states is
considerably reduced in comparison to the F1--F2~distance in
opposite states, and the hole--hole repulsion energy varies
accordingly. Of course, one has to compare states which have a
large F1$^{-1}$F2$^{-1}$~population that arises from the same
types of orbitals. Such F1$^{-1}$F2$^{-1}$~states are
distributed over a small energy range in the double ionization spectrum.
Corresponding lines in~\XeFt\ mark the lower ends of such
DIP ranges, due to the maximum distance between the two vacancies.
Coulomb repulsion between holes on adjacent fluorine ligands 
in~\XeFf\ and \XeFs\ results in shifts toward higher DIPs.

States with a large Xe$^{-1}$F$^{-1}$~population do not group like
those of F1$^{-1}$F2$^{-1}$~character. The density of these
Xe$^{-1}$F$^{-1}$~states also increases with an increasing number
of fluorine atoms, but they are distributed more uniformly over the
whole spectral range. In~\XeFt\ and~\XeFs, these states are
dominant with respect to population number, and in~\XeFf\ they are
comparable to the F1$^{-1}$F2$^{-1}$~states. The~DIPs of the
Xe$^{-1}$F$^{-1}$~states are not subject to a change of the
hole--hole repulsion energy (in contrast to those of the
F1$^{-1}$F2$^{-1}$~states) because the Xe--F~distance is the same
for all fluorine ligands within a molecule (remember that the double 
ionization spectrum of~\XeFs\ was calculated assuming octahedral symmetry).

\subsection{Xe\hspace{1.66672pt}4\textit{d} Linewidth}
\label{sec:Xe4d-line}

The electronic decay processes presented in Sec.~\ref{sec:decayproc} 
are characterized in Sec.~\ref{sec:widths} according to their
final-state population. Using Eqs.~(\ref{eq:partial}), (\ref{eq:defs1}),
and (\ref{eq:defs2}), the analysis of the final-state population numbers 
allows one to determine the relative importance of IAED, ICD, ETMD2, and ETMD3 
for the electronic decay of a Xe$\,4d$~hole in \XeFn. The partial decay 
width associated with a given decay process can be viewed, according to 
Eq.~(\ref{eq:defs1}), as a product of an averaged Coulomb matrix element 
and a two-hole population factor related to that process. 

\begin{table}
  \centering
  \begin{tabular}{c|c|c|c|c}
    Compound & $Q\I{IAED}$ & $Q\I{ICD}$ & $Q\I{ETMD2}$ & $Q\I{ETMD3}$ \\
    \hline\hline
          Xe & 14.6 & 0    & 0    & 0      \\
    \hline
       \XeFt & 10.4 & 43.8 & 19.4 &  \phantom{0}22.7 \\
    \hline
       \XeFf & \phantom{0}7.1 & 68.1 & 34.3 & 124.1 \\
    \hline
       \XeFs & \phantom{0}4.8 & 80.2 & 45.5 & 288.6 \\
  \end{tabular}
  \caption{Two-hole population factors for intra-atomic electronic decay 
           (IAED) and for the interatomic decay processes ICD, ETMD2, 
           and ETMD3 in xenon and its fluorides. The two-hole population 
           factors are defined in Eq.~(\ref{eq:defs2}).}
  \label{tab:pop}
\end{table}

We calculated the population factors, as defined by Eq.~(\ref{eq:defs2}), by 
summing population numbers up to the energy of the Xe$\,4d$~lines obtained 
in Ref.~\onlinecite{Buth:-03}. The results are collected in 
Table~\ref{tab:pop}. We see that~$Q\I{IAED}$ is rather low in the xenon 
fluorides, especially when compared to the population factors~$Q\I{ICD}$, 
$Q\I{ETMD2}$, and $Q\I{ETMD3}$. Hence, as far as 
population numbers are concerned, IAED is suppressed, and interatomic 
decay processes dominate the electronic decay of a Xe$\,4d$~hole. 
Also note the sensitive dependence of all population factors on the number
of fluorine ligands.

In order to arrive at a complete picture, the transition strengths
in Eq.~(\ref{eq:defs1}) have to be determined. For this purpose, we make
a reasonable assumption: The transition strengths~$|T\I{IAED}|^2$,  
$|T\I{ICD}|^2$, $|T\I{ETMD2}|^2$, and $|T\I{ETMD3}|^2$ are universal for xenon
and its fluorides. This allows us---utilizing Eqs.~(\ref{eq:partial}) and
(\ref{eq:defs1})---to take the averaged experimental Xe$\,4d$~linewidths, 
displayed in Fig.~\ref{fig:Gamma_XeFx} and reproduced in 
Table~\ref{tab:widths}, and the population factors in Table~\ref{tab:pop} 
to set up a linear system of four equations for the four unknown 
transition strengths. The matrix of population factors, however, is extremely
ill-conditioned, as its $\mathbb L^2$~condition number,~\cite{Golub:MC-89} for 
instance, is~$945$, which is much larger than unity. This means that the 
solution of the linear system depends strongly on the accuracy of the 
population factors. 

Therefore, we first determine~$|T\I{IAED}|^2$ by solving 
Eqs.~(\ref{eq:partial}) and (\ref{eq:defs1}) for the xenon atom. 
We then exploit that both~$|T\I{ETMD2}|^2$ and $|T\I{ETMD3}|^2$ are based 
on electron transfer between Xe and a fluorine ligand. They are thus expected 
to be of similar magnitude, so we set~$|T\I{ETMD2}|^2 = |T\I{ETMD3}|^2$. The 
remaining linear system of three equations and two unknowns can be solved in 
a stable fashion by a least-squares procedure.~\cite{Golub:MC-89} The 
calculated transition strengths for~Xe and \XeFn\ are displayed in 
Table~\ref{tab:Tfactors}. The results are in line with expectations: The 
transition strength associated with intra-atomic decay is greater by an 
order of magnitude than the interatomic energy-transfer strength of ICD.
$|T\I{ETMD2}|^2$ and $|T\I{ETMD3}|^2$, which require interatomic orbital
overlap for electron transfer, are even smaller.

\begin{table}
  \centering
  \begin{tabular}{c|c}
    Decay Process & $|T|^2$ \ \rm [eV]\\
    \hline\hline
    IAED & $1.4\E{-2}$ \\
    \hline
    ICD          & $1.9\E{-3}$ \\
    \hline
    ETMD2/ETMD3  & $2.1\E{-4}$ \\
  \end{tabular}
  \caption{Transition strengths for electronic decay of a Xe$\,4d$~hole
           in \XeFn.}
  \label{tab:Tfactors}
\end{table}

\begin{table*}
  \centering
  \begin{tabular}{c|c|c|c|c|c}
    Compound & $\Gamma\I{IAED}  \rm\ [eV]$ & $\Gamma\I{ICD}   \rm\ [eV]$ & 
               $\Gamma\I{ETMD2} \rm\ [eV]$ & $\Gamma\I{ETMD3} \rm\ [eV]$ & 
               $\Gamma\I{expt}  \rm\ [eV]$ \\
    \hline\hline
	  Xe & 0.20 & \phantom{00}0 & \phantom{00}0 & \phantom{00}0 & 0.20 \\
    \hline
       \XeFt & 0.15 & 0.08 & 0.00 & 0.00 & 0.24 \\
    \hline
       \XeFf & 0.10 & 0.13 & 0.01 & 0.03 & 0.26 \\
    \hline
       \XeFs & 0.07 & 0.15 & 0.01 & 0.06 & 0.29 \\
  \end{tabular}
  \caption{Decomposition of the mean experimental electronic decay width,
          $\Gamma\I{expt}$, of the Xe$\,4d$-lines in~Xe and \XeFn\ into 
          contributions from intra-atomic and interatomic decay processes.
          $\Gamma\I{expt}$ is taken from Fig.~\ref{fig:Gamma_XeFx}.}
  \label{tab:widths}
\end{table*}

Finally, by combining the population factors from Table~\ref{tab:pop} and the
transition strengths from Table~\ref{tab:Tfactors}, the partial 
widths~$\Gamma\I{IAED}$, $\Gamma\I{ICD}$, $\Gamma\I{ETMD2}$, and $\Gamma\I{ETMD3}$
in~\XeFn\ can be computed [Eq.~(\ref{eq:defs1})]. They are presented in 
Table~\ref{tab:widths}. The conclusions that can be drawn from this table are
truly remarkable. From~Xe to~\XeFs, the contribution from~IAED to the total 
Xe$\,4d$~decay width drops monotonically, consistent with the one-center
picture mentioned in Sec.~\ref{sec:intro}. However, that decrease is more than
compensated for by interatomic decay, which is the reason why an overall increase
in Xe$\,4d$~width is observed in experiment. ICD~is the most important 
interatomic decay mechanism. In~\XeFf, $\Gamma\I{ICD}$ is already greater than  
$\Gamma\I{IAED}$, and in~\XeFs, more than~$50 \, \%$ of the total Xe$\,4d$~width
are caused by ICD. Decay via electron transfer is less significant. 
Nevertheless, while ETMD2 is practically negligible throughout, $\Gamma\I{ETMD3}$
and $\Gamma\I{IAED}$ in~\XeFs\ are in fact comparable, due to the steep rise of
the population factor~$Q\I{ETMD3}$ between~\XeFt\ and \XeFs\ (Table~\ref{tab:pop}).

\section{Conclusion}
\label{sec:conclusion}

We have shown that interatomic electronic decay processes are of prime importance
for understanding relaxation of a Xe$\,4d$~vacancy in the xenon fluorides. The 
experimentally observed trend of increasing Auger width in the series~\XeFn, 
$n=0,2,4,6$, is an impressive consequence of the fact that interatomic
Coulombic decay and electron-transfer mediated decay not only cancel out the drop 
in intra-atomic Auger rate: Interatomic decay dominates by far in the larger xenon 
fluorides. In~\XeFs, for example, $76 \, \%$~of the total Xe$\,4d$~linewidth is due 
to the combined effect of~ICD, ETMD2, and ETMD3.

ICD in~\XeFn\ takes place on an ultrashort time scale of less than 10~fs. This is 
of the same order of magnitude as the ICD lifetimes calculated for a $2s$~hole in 
a Ne atom surrounded by at least six neon monomers.~\cite{Santra:ED-01} Our 
finding that, in~\XeFs, $\Gamma\I{ICD}$ is only a little more than two times
greater than~$\Gamma\I{ETMD3}$, affords a new impulse for cluster and 
condensed-matter research. We anticipate ETMD3 to be a major decay mechanism for
an inner-valence hole, for instance, in a neon atom embedded in an argon matrix.

There are obvious consequences of our investigation for molecules bearing a 
similarity to~\XeFn. An interesting example is Auger decay of a Si$\,2p$~hole in
SiF$_4$. It was shown in Refs.~\onlinecite{Tarantelli:FI-93} and 
\onlinecite{Gottfried:FI-96} that virtually not a single dicationic (valence) 
eigenstate in SiF$_4$ is, at a significant level, of Si$^{-2}$~character. The 
occurrence of primarily multi-center decay channels is termed \emph{foreign 
imaging}. It is these channels to which~ICD, ETMD2, and ETMD3 provide access. In 
light of the present study, we may conclude that the calculation by 
Larkins,~\cite{Larkins:CHS-94} based on the one-center model, must be
expected to appreciably underestimate the Si$\,2p$~Auger width in SiF$_4$. A recent
measurement by Thomas~\etal~\cite{Thomas:ANL-02} proves this expectation 
to be justified: The experimental Si$\,2p$~Auger width is more than five times
greater than the prediction of the one-center model. The authors of 
Ref.~\onlinecite{Thomas:ANL-02} also interpret this discrepancy in terms of
interatomic electronic processes.

Our approach to molecular Auger decay, as presented in this paper, lays its 
emphasis on the decay channels. Though not impossible, an actual computation
of Auger widths in such complicated many-electron systems as the xenon fluorides
still cannot be carried out in a very reliable and accurate manner. We therefore
put to use two highly powerful theoretical tools: the ADC scheme for the 
two-particle propagator and the two-hole population analysis of dicationic eigenstates.
These two methods allow us to concentrate on the mechanisms underlying Auger decay
by yielding detailed information about the nature of the available decay channels.

The key relations~(\ref{eq:partial}), (\ref{eq:defs1}), and (\ref{eq:defs2}) between
two-hole population numbers and partial Auger widths are, strictly speaking, only
valid for molecules with pronounced charge localization characteristics. The
xenon fluorides satisfy this criterion well. We are convinced, however, that
the basic picture is much more general. In order to understand Auger decay in
arbitrary molecules, it is usually insufficient to restrict the analysis to local
valence electron densities. The physics may be even more complicated, but a 
qualitative decomposition into interatomic processes~ICD, ETMD2, and ETMD3 is most
likely still possible.

\begin{acknowledgments}
The authors are highly indebted to T. Darrah Thomas for pointing
out Ref.~\onlinecite{Cutler:Xe-91} and supporting it further with 
valuable private communications (Fig.~\ref{fig:Gamma_XeFx}). 
This work would not have been possible without the ADC~programs
and support by Francesco Tarantelli. Imke B. M\"uller, Sven
Feuerbacher, J\"org Breidbach, and Thomas Sommerfeld accompanied
our work with helpful comments and fruitful discussions. R.S.~and
L.S.C. gratefully acknowledge financial support by the Deutsche
Forschungsgemeinschaft~(DFG).
\end{acknowledgments}

\end{document}